\documentclass[%
 reprint,
superscriptaddress,
 amsmath,amssymb,
 aps,
prb,
]{revtex4-1}

\usepackage{xcolor}

\usepackage{graphicx}% Include figure files
\usepackage{dcolumn}% Align table columns on decimal point
\usepackage{bm}% bold math

\def\be{\begin{eqnarray}}
\def\ee{\end{eqnarray}}
\def\ii{{\mathrm i}}

\def\r{{\bm r}}

\def\E{{\bm E}}
\def\H{{\bm H}}

\def\G{{\bf G}}
\def\GL{\mathbb{G} }
\newcommand{\alphag}{\bm{\alpha}}

\def\II{{\overleftrightarrow{{\bf I}}}}
\def\GG{{\overleftrightarrow{{\bf G}}}}
\def\GGL{{\overleftrightarrow{{\mathbb{G} }}}}
\def\alphagg{{\overleftrightarrow{{\bm{\alpha}}}}}
\newcommand{\alphaggt}{{\widetilde
{\overleftrightarrow{\bm{\mathbb{\alpha}}}}}}

\begin{document}

\title{Coupled electric and magnetic dipole formulation for planar arrays of 
%dipolar 
particles: resonances and bound states in the continuum for all-dielectric metasurfaces 
%with various electric and/or magnetic meta-atoms per unit cell
}

\author{Diego R. Abujetas}

\email{diego.romero@iem.cfmac.csic.es}
\affiliation{%
 Instituto de Estructura de la Materia (IEM-CSIC), Consejo Superior de Investigaciones Cient\'{\i}ficas,\\ Serrano 121, 28006 Madrid, Spain
}%
\author{Jorge Olmos-Trigo}
\author{Juan J. S\'aenz}%
\affiliation{Donostia International Physics Center DIPC, Paseo Manuel de Lardizabal 4, 20018, Donostia, San Sebastián, Spain}

\author{Jos\'e A. S\'anchez-Gil}%
 \email{j.sanchez@csic.es}
\affiliation{%
 Instituto de Estructura de la Materia (IEM-CSIC), Consejo Superior de Investigaciones Cient\'{\i}ficas,\\ Serrano 121, 28006 Madrid, Spain
}%
%\affiliation{Donostia International Physics Center DIPC, Paseo Manuel de Lardizabal 4, 20018, Donostia, San Sebastián, Spain}

\date{\today}

\begin{abstract}
The optical properties of infinite planar array of scattering particles, metasurfaces and metagratings, are attracting special attention lately for their rich phenomenology, including both plasmonic and high-refractive-index dielectric meta-atoms with a variety of electric and magnetic resonant responses. Herein we derive a coupled electric and magnetic dipole analytical formulation to describe the reflection and transmission of such periodic arrays, including specular and diffractive orders, valid in the spectral regimes where only dipolar multipoles are needed. 
%Electric and/or magnetic dipoles with all three orientations arising in turn from a single or various meta-atoms per unit cell are considered. 
The 2D lattice Green function is rewritten in terms of a 1D (chain) version that fully converges in the complex frequency plane and can be easily calculated. Modes emerging as poles of such lattice Green function can be extracted, as evidenced by calculating resonances and bound states in the continuum for an array of Si spheres. This formulation can be applied to investigate a wealth of plasmonic, all-dielectric, and hybrid metasurfaces/metagratings of interest throughout the electromagnetic spectrum.

\end{abstract}

%\keywords{Suggested keywords}%Use showkeys class option if keyword
                              %display desired
\maketitle

\section{Introduction} 

Planar arrays of  resonant particles are attracting a great deal of attention nowadays. Widely known as metasurfaces (sub-wavelength lattice constant, thus in the non-diffraction spectral regime) and metagratings (diffraction being relevant), they exhibit fascinating properties that indeed hold promise of infinitely thin optical devices performing a wealth of  functionalities \cite{Holloway2012a,Glybovski2016,Haltout2016,Li2017c,Genevet2017,Neshev2018,Qiao2018,Sun2019,Kupriianov2019,Paniagua-Dominguez2019,Staude2019,Sain2019a,Pertsch2020}. Although consisting in the beginning of metallic meta-atoms \cite{Abajo2005,GarciadeAbajo2007,Meinzer2014,Genevet2017}  typically supporting localized plasmons (mostly electric dipole) resonances, they have been extended in recent years to include high-refractive-index (HRI) particles which possess Mie resonances with a (lowest-order) magnetic dipole character \cite{Evlyukhin2010,Garcia-Etxarri2011,Paniagua-Dominguez2011,Nieto-Vesperinas2011,Kuznetsov2012,Kuznetsov2016,Zhu2016}.

Furthermore, the unconventional optical properties of metasurfaces stem not only from the  resonant properties of the meta-atoms themselves, but also from multiple scattering effects through coupling with guided-mode or lattice resonances~\cite{Laroche2006,Gomez-Medina2006,Marinchio2014,Holloway2012,Ko2018,Tagay2018,Abujetas2018a,Abujetas2019a}. Dealing with such complex interactions between particle and  lattice resonances in planar infinite arrays by means of full wave numerical calculations is a formidable tasks even for common available solvers, which in turn cannot shed much light onto the underlying physics.  In this regard, coupled-dipole formulations have been developed since long ago ~\cite{Purcell1973,Bohren_Huffman}, 
extended to a wealth of configurations typically involving a finite number of dipoles. 
The widely employed discrete dipole approximation  is an extension of such coupled-dipole formulations  to deal with macroscopic objects discretized through volume (dipolar) elements~\cite{Draine1994,Yurkin2007}. 
Nonetheless, very few works have  investigated  infinite dipolar arrays thoroughly, especially for 2D planar arrays involving both electric and magnetic dipole resonances~\cite{Evlyukhin2010,Sersic2012,Swiecicki2017a,Swiecicki2017,Babicheva2017a,Chen2017,Baur2018,Babicheva2019,Abujetas2019d,Abujetas2019c}, the latter crucial to account for lowest-order Mie resonances of HRI particles. 

In this regard, a formulation that rigorously addresses  lattice interactions is paramount to properly describe collective resonances and especially  bound states in the continuum with diverging Q factors in metasurfaces \cite{Marinica2008,Hsu2016a,Koshelev2018,Minkov2018,Koshelev2019a,Kupriianov2019,Abujetas2019d,Abujetas2019c}. Metasurfaces operating at/near such resonances hold indeed promise of a variety of planar devices with functionalities such as sensing \cite{Yanik2011,Yesilkoy2019}, filtering \cite{Foley2014}, lasing \cite{Kodigala2017,Ha2018,Khaidarov2019,Wu2020}, and enhanced non-linearities \cite{Carletti2018}.

In this work we analyze analytically the reflection and transmission of two-dimensional lattices of one or various meta-atoms supporting electric and/or magnetic dipolar resonances with arbitrary orientation in- and out-of-plane. We develop a coupled electric and magnetic dipole (CEMD) formulation that fully accounts for the coupling between those electric and magnetic dipolar fields. In fact, this CEMD formulation has been successfully exploited to deal with HRI disk metasurfaces~\cite{Abujetas2019d} and rod dimer arrays~\cite{Abujetas2019c}, but most details remain unpublished. Here we also incorporate a significant improvement, pertaining to the calculation of the so-called 2D lattice Green function (from which all relevant magnitudes are derived): rather than extracting it numerically through convergence, we rewrite it in terms of a 1D lattice Green function that can be easily evaluated. Finally, we also analyze the emergence of various lattice modes from the poles of the thus obtained Green function.

\section{Formulation of the scattering problem}

Let us consider an infinite set of identical particles arranged in a rectangular array. Without loss of generality, the particle labeled as $(n,m)$ is placed at
\begin{equation}
\r_{nm} = x_n \hat{x} + y_n \hat{y} = na \hat{x} + mb \hat{y},
\end{equation}
where $a$ and $b$ are the lattice constants along the $x$ and $y$ axis, respectively. The array is then illuminated by an external plane wave, $\bm{\Psi}^{(0)}(\r)$, with incident wavevector $\mathbf{k^{(0)}} = k_x^{(0)}\hat{x} + k_y^{(0)}\hat{y} + k_z^{(0)}\hat{z}$, while the time dependence
$\exp\left(-i\omega t\right)$ will be assumed for all the fields;  $\omega$ is the angular frequency, related to the modulus of the wavevector through $k = \omega/c$, $c$ being the speed of light. In what follows, $\bm{\Psi}(\r)$ stands for the electromagnetic field components needed to fully describe the scattering problem, namely, the electric and/or magnetic fields and/or any combination of both; thus its representation depends on the basis.

Each particle in the array is excited by the external plane wave plus the waves scattered from the rest of the array
%\cite{Twersky1952}. 
The self-consistent incident field, $\bm{\Psi}_{\text{inc}}(\r)$, on the $(n,m) = (0,0)$ particle ($\r_{00} =  \bm 0$),
 is then given by the solution of %r_{00}
\be
\bm{\Psi}_{\text{inc}}(\bm 0) 
=
\bm{\Psi}^{(0)}(\bm 0) 
+
\sum_{nm}{'}
k^2 \GG(-\r_{nm}) \alphagg
\bm{\Psi}_{\text{inc}}(\r_{nm}),
\ee
where $\sum_{nm}{'}$ means that the sum runs for all indices except for $(n, m) = (0, 0)$. $\GG(\r)$ and $\alphagg$ are matrices representing the dyadic Green function and the dipolar polarizability of the particles, respectively,  and their representations depend on the chosen basis to describe the electromagnetic fields, $\bm{\Psi}(\r)$. The dyadic Green function is obtained from the scalar Green function, $g\left(\r\right)$, by applying a linear differential operator, $\mathbf{\mathcal{L}}$, that also depends on the chosen basis (see below).

For a periodic array and plane wave illumination the Bloch's theorem holds, $\bm{\Psi}_{\text{inc}}(\r_{nm})  = \bm{\Psi}_{\text{inc}}(\bm 0) \exp(\ii k_x^{(0)} n a)\exp(\ii k_y^{(0)} m b) = \bm{\Psi}_{\text{inc}}(0) e^{\ii\phi _{nm}}$, and the self-consistent incident field can be written as
\begin{align}
\bm{\Psi}_{\text{inc}}(\bm 0)
&=
\bm{\Psi}^{(0)}(\bm 0) + k^2\left[\sum_{nm}{'} \GG(-\r_{nm}) e^{\ii\phi_{nm}}\right] \alphagg \bm{\Psi}_{\text{inc}}(\bm 0) \nonumber \\
 &\equiv \bm{\Psi}^{(0)}(\bm 0) + k^2 \GGL_{b}\alphagg \bm{\Psi}_{\text{inc}}(\bm 0).
\label{Eq:int}
\end{align}
We have defined $\GGL_{b}$, the lattice `depolarization' dyadic (or return Green function), as 
\be
\GGL_{b} \equiv \sum_{nm}{'}
\GG(-\r_{nm})e^{\ii\phi _{nm}}.
\label{eq:Gb_2D_ini}
\ee
$\GGL_{b}$ tells us about the coupling strength between particles, and is crucial to determine all the lattice properties.  Next, the solution of the self-consistent equation can be formally written as a function of the external plane wave as:
\begin{equation}
\bm{\Psi}_{\text{inc}}(\bm 0)
= 
\left[\II - k^2 \GGL_{b}\alphagg \right] ^{-1}
\bm{\Psi}^{(0)}(\bm 0),
\label{eq:cemd}
\end{equation}
where $\II$ is the unit dyadic.

Once we know the self-consistent incoming field, Eq.~\eqref{eq:cemd}, the field scattered by the $(n,m)$ particle is given by
\be
\bm{\Psi}_{\text{scat}}(\r) &=&  k^2 \GG(\r-\r_{nm})  \alphagg \ \bm{\Psi}_{\text{inc}}(\r_{nm}),
 \ee
and the total scattered field can be written as  
\be
\bm{\Psi}_{\text{scat-tot}}(\r)
&=&
k^2 \left\{\sum_{nm}
 \GG(\r-\r_{nm}) e^{\ii\phi _{nm}}
\right\} \alphagg  \bm{\Psi}_{\text{inc}}(\bm 0) \nonumber \\ &\equiv& k^2 \GGL^\pm(\r) \alphagg  \bm{\Psi}_{\text{inc}}(\bm 0),  \label{eq:Etot0}
\ee 
where the tensor lattice sum $\GGL^{\pm}(\r)$ can be written as a sum over all diffracted spectral orders  ($l,p = \cdots, -2, -1, 0, 1, 2 \cdots$)
 as:
 \be
\GGL^{\pm}(\r)  &\equiv&  \sum_{nm}
 \GG(\r-\r_{nm})  e^{\ii\phi_{nm}} \nonumber \\ &=& \sum_{l,p} \GGL^{\pm}_{lp} e^{\ii k_x^{(l)}x} e^{\ii k_y^{(p)}y} e^{\pm \ii k_z^{(l,p)} z} \nonumber \\
  &=& \sum_{l,p} \GGL^{\pm}_{lp} e^{\pm\ii\phi_{lp}(\r)},
\label{Eq:TGF} 
\ee
where, for $k_z^{(0)} > 0$, ``+'' (``-'') corresponds to upward scattered waves in the region $z>0$ (downwards reflected waves in the region $z<0$). $k_x^{(l)}$,  $k_y^{(p)}$ and $k_z^{({l,p})}$ are the wavevectors of the diffracted orders
\be
k_x^{(l)} = k_x^{(0)} - \dfrac{2\pi}{a}l, \quad k_y^{(p)} = k_y^{(0)} - \dfrac{2\pi}{b}p, \nonumber \\
k_z^{({l,p})} = \sqrt{k^2 - (k_x^{(l)})^2 - (k_y^{(p)})^2}.
\ee
Similar to the dyadic Green function, the tensor lattice sum can be written as the differential operator $\mathbf{\mathcal{L}}$ applied to a scalar quantity
\be
\sum_{l,p}\GGL_{lp}^{\pm}e^{\pm\ii\phi_{lp}(\r)} =  \sum_{l,p}\dfrac{\ii}{2abk_{z}^{(lp)}} \mathbf{\mathcal{L}} e^{\pm\ii\phi_{lp}(\r)}.
\ee
%See Eq. \eqref{GGLcomcar} in Appendix \ref{Glattice} for explicit expressions.
Finally, using Eq.~\eqref{eq:cemd} in Eq.~\eqref{eq:Etot0}, the reflected and transmitted fields are then given by
 \be \bm{\Psi}_{\text{r}}(\r)  &=&   k^2 \GGL^-(\r)  \alphaggt \ \bm{\Psi}^{(0)}(\bm 0), \label{totref} \\
 \bm{\Psi}_{\text{t}}(\r)  &=& \bm{\Psi}^{(0)}(\r) +  k^2 \GGL^+(\r)  \alphaggt \ \bm{\Psi}^{(0)}(\bm 0), \label{tottrans}
\ee
 where $\alphaggt$ is the renormalized (dressed) polarizability,
\begin{equation}
   \alphaggt = \alphagg \left[\II - k^{2}\GGL_{b} \alphagg \right]^{-1} = \left[\alphagg^{-1} - k^{2}\GGL_{b} \right]^{-1}. 
\label{eq:alpharnorm}
\end{equation}
Writing $\GGL^{\pm}(\r)$ as a sum over diffracted spectral orders, the field scattered into each diffractive mode is
 \be \bm{\Psi}_{\text{r}}^{(l,p)}(\r)  &=&   k^2 \GGL_{lp}^-  \alphaggt \ \bm{\Psi}^{(0)}(\bm 0) e^{-\ii\phi_{lp}(\r)}, \label{totref_lp} \\
 \bm{\Psi}_{\text{t}}^{(l,p)}(\r)  &=& k^2 \GGL_{lp}^+  \alphaggt \ \bm{\Psi}^{(0)}(\bm 0) e^{+\ii\phi_{lp}(\r)}. \label{tottrans_lp}
\ee

\section{Lattice depolarization Green function for arbitrary 2D arrays}

The optical properties of periodic arrays are then described by the lattice depolarization Green function, $\GGL_b$, that accounts for the electromagnetic field scattered by all the array over its own particles. The evaluation of $\GGL_b$ can be done in real space, but the convergence is in general very low. Although there are mathematical techniques to improve the convergence, ~\cite{Stillinger1990}, 
it is more convenient to transform the sum from the real to the reciprocal space. For example, for complex frequencies the sum cannot be evaluated in the real space. Actually, the (complex) resonant frequencies of the metasurface can be only found in the reciprocal space, which in turn yields more physical insights and approximate expressions close to the Rayleigh-Wood anomalies.
To simplify the expressions, the sum for a 2D array of particles will be described below as that for a 1D array, namely, a chain of particles. To this end, let us first study the scattering properties of a chain of particles.

\subsection{Scattered field by a chain of particles}

The scattering properties of an individual particle are derived from the scalar Green function, $g\left(\r - \r'\right)$, defined as the solution of the Helmholtz equation with a point source located at $\r=\r'$:
\be
\nabla^2 g\left(\r - \r' \right) + k^2 g\left(\r - \r' \right) = - \delta\left(\r - \r' \right),
\ee
where $\r$ is the observation point, $\r'$ is the position of the emitter/source, and $k$ is the wavevector in the media. The scalar Green function in 3D is a spherical wave that propagates away from its origin
\begin{align}
g\left(\r - \r' \right) & = \dfrac{e^{\ii k |\r - \r'|}}{4\pi |\r - \r'|} \nonumber \\ & =
\int \dfrac{\mathrm{d}Q_x \mathrm{d}Q_y}{4\pi^2} e^{\ii Q_x \left(x - x' \right)}e^{\ii Q_y \left(y - y' \right)} \dfrac{\ii}{2q} e^{\ii q |z - z' |}, \nonumber \\ & \quad \quad \quad \quad \quad \quad q = \sqrt{k^2 - Q_x^2 - Q_y^2},
\label{eq:GF3D}
\end{align}
wherein we have used the Weyl expansion to express it as a sum of plane waves. The dyadic Green function, $\GG\left(\r - \r' \right)$, is obtained from the scalar Green function as $\mathbf{\mathcal{L}}g\left(\r - \r' \right)$ 
%by applying a linear differential operator, $\mathbf{\mathcal{L}}$, that depends on the basis chosen to describe the electromagnetic fields.

Now let us consider a periodic chain of particles along the $x$ axis located at the position $\r_n$
\begin{equation}
\r_n = x_n \hat{x} = na \hat{x},
\end{equation}
$a$ being the separation between particles.
Upon the incidence of an external plane wave (with incident wavevector, $\mathbf{k^{(0)}} = k_x^{(0)}\hat{x} + k_y^{(0)}\hat{y} + k_z^{(0)}\hat{z}$), the scattering properties of the chain of particles are defined by the next sum in the real space (analogous to $\GGL^{\pm}(\r)$ for 2D arrays):
\begin{equation}
\sum_{n} \G\left(\r - \r_n \right) e^{\ii k_x^{(0)} n a} = \mathbf{\mathcal{L}}\sum_{n} g\left(\r - \r_n \right) e^{\ii k_x^{(0)} n a},
\label{eq:gl}
\end{equation}
where $k_x^{(0)}$ is the projection of the incident wavevector along the $x$ axis. The linear differential operator is made of spatial derivatives, independent of the summation variable $n$, so we place it out of the summation in the latter equation.  %(The incident plane is the $xz$ plane). 
After some algebraic manipulations (using the Weyl representation), the sum in the real space can be rewritten in the reciprocal space as:
\begin{equation}
\sum_{n} g\left(\r - \r_n \right) e^{\ii k_x^{(0)} n a} = \sum_{l} e^{\ii k_x^{(l)}x} \dfrac{\ii}{4a} H_0 \left(k_{||}^{(l)} \rho \right),  
\label{eq:5}
\end{equation}
where $H_0$ is the 0th-order Hankel function of the first type with $\rho = \left(y^2 + z^2 \right)^{1/2}$ in its argument being the radial distance to the chain of particles; $k_{||}^{(l)}$ and $k_x^{(l)}$ are the wavevector components of the diffracted waves
\begin{equation}
k_{||}^{(l)} = \sqrt{k^2 - \left(k_x^{(l)}\right)^2}, \quad k_x^{(l)} = k_x^{(0)} - \dfrac{2\pi}{a}l.
\end{equation}
The scalar Green function in 2D with translational symmetry along the $x$ axis (i.e., for an infinitely long cylinder where the $x$ axis is the cylinder axis) is 
\begin{equation}
g_{2D}\left(\r - \r_n \right) = \dfrac{\ii}{4} H_{0}\left(\sqrt{k^2 - k_x^2}|\r - \r_n|\right) e^{\ii k_{x} x}.
\label{eq:gb2D}
\end{equation}
Thus, Eq.~\eqref{eq:5} can be written as:
\begin{equation}
\sum_{n} g\left(k,\r - \r_n \right) e^{\ii k_x^{(0)} n a} = \sum_{l} \dfrac{1}{a} g_{2D}\left(k,k_x^{(l)},\r - \r_n \right),
\label{eq:9}
\end{equation}
where we have explicitly added the wavevector as an argument in the scalar Green functions (normally this is omitted). 

Therefore, from Eq.~\eqref{eq:9} we can infer that a chain of particles behaves effectively as an infinitely long cylinder that is however  excited by both propagating and evanescent plane waves. There is always a propagating wave that coincides with an incident external plane wave, while the rest of propagating waves appears as a diffraction phenomenology when the lattice constant $a$ exceeds the incoming wavelength $\lambda = 2\pi/k$. Thus, only terms with $\mathrm{Im}\left[k_{||}^{(l)}\right] = 0$ survive in the far field , whereas in the near field there are contributions from all waves.

The other key element to describe the optical response of a chain of particles is the lattice depolarization Green function, that is defined for this system as
\be
\GGL_{b-Ch}(a,k,k_{x}^{(0)}) = \mathbf{\mathcal{L}}\sum_{n}{'} g\left(\r - \r_n \right) e^{\ii k_x^{(0)} n a}.
\label{eq:Gbch_l}
\ee
Importantly, $\GGL_{b-Ch}$ can be expressed in terms of analytical functions called polylogarithm functions. 
%(Appendix?)

\subsection{Rectangular arrays}

Now we have the tools to tackle the problem of the two dimensional rectangular array of particles. By using $\mathbf{\mathcal{L}}$, Eq.~\eqref{eq:Gb_2D_ini} can be rewritten as:
\be
\GGL_b \equiv && \sum_{nm}{'}
\GG(-\r_{nm})e^{\ii\phi _{nm}} \nonumber \\ && = \mathbf{\mathcal{L}} \sum_{nm}{'} g(-\r_{nm})e^{\ii\phi _{nm}}.
\ee
From Eq.~\eqref{eq:9} we learn that a 2D array of particles can be seen as a 1D array of chains of particles. Assuming that the chains are along the $x$ axis, the effective cylinders (that the chains represent) are placed at $\r_{m} = mb \hat{y}$. Within this convention, it is straightforward to show that 
\be
 \GGL_{b} =  \GGL_{b-Ch} +  \dfrac{1}{a}\left( \sum\limits_{l} \GGL_{b-1D}^{(l)} \right),
 \label{eq:Gb2D}
\ee 
where $\GGL_{b-Ch}$ is the `depolarization' dyadic of a chain of particles defined in Eq.~\eqref{eq:Gbch_l}
and $\GGL_{b-1D}^{(l)}$ is the `depolarization' dyadic of a 1D array of cylinders (with their axis along the $x$ axis):
\be
&& \GGL_{b-1D}^{(l)}(b,k,k_{y}^{(0)},k_x^{(l)},\r) \nonumber \\ && \hspace{1.5cm} = \mathbf{\mathcal{L}} \sum_{m}{'} g_{2D}(k,k_x^{(l)},-\r_{m}) e^{\ii k_y^{(0)} m b}.
\label{eq:Gb1D_l}
\ee
In both cases all arguments of the functions are shown explicitly. The importance of writing $\GGL_{b}$ as shown in Eq.~\eqref{eq:Gb2D} is that analytical expressions are available for Eq.~\eqref{eq:Gbch_l} and  Eq.~\eqref{eq:Gb1D_l} in the reciprocal, although some care must be taken regarding the convergence of the expressions. For large $|k_x^{(l)}| \gg k$, the convergence of Eq.~\eqref{eq:Gb1D_l} worsens. Thus, if $m^{(l)}$ is defined as the number of terms needed to achieve convergence, $m^{(l)}$ will increase with increasing $|k_x^{(l)}|$. Fortunately, Eq.~\eqref{eq:Gb2D} presents a good convergence with respect to the index $l$ ($\GGL_{b-1D}^{(l)}$ goes rapidly to zero as $l$ increases), and only a few orders are needed to achieve accurate results.

\begin{figure}
\includegraphics[width=1\columnwidth]{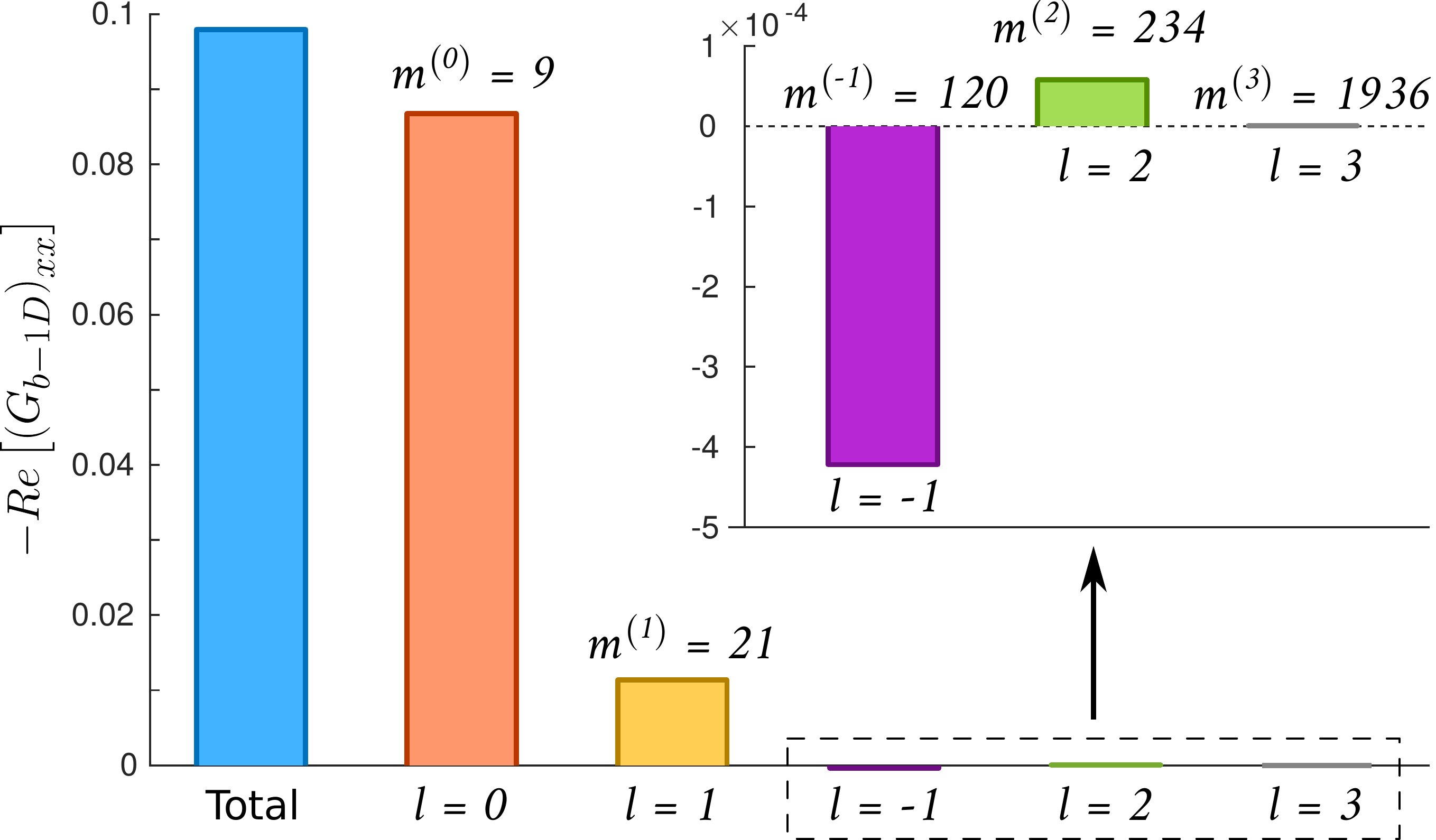}
\caption{Real part of the $xx$ matrix term in Eq.~\eqref{eq:Gb1D_l} for different values of $l$ at $kb = \pi$, $k_x^{(0)} = 0.7k\cos(\alpha)$, $k_y^{(0)} = 0.7k\sin(\alpha)$, and $\alpha = 10^{\circ}$. Also, the value of $m^{(l)}$ is indicated for each diffractive order.}
\label{fig_conver}
\end{figure}

For the sake of clarity, we plot in Fig.~\ref{fig_conver} the numerical values of the $xx$ matrix term (the other terms present a similar behaviour) of $\GGL_{b-1D}^{(l)}$ [see Eq.~(A2)] for an arbitrary set of parameters: we choose $kb = \pi$, $k_x^{(0)} = 0.7k\cos(\alpha)$, $k_y^{(0)} = 0.7k\sin(\alpha)$, and $\alpha = 10^{\circ}$. Only the real part is plotted because the convergence of the imaginary part is much faster. In fact, a diffractive order $l$ at real frequencies only contributes to the imaginary part if $\mathrm{Im}\left[k_{||}^{(l)}\right] = 0$. This is important because the imaginary part of the depolarization Green function is directly related to the optical theorem. Therefore, inaccurate calculations of such imaginary part can lead to nonphysical results in the calculation of the reflectance, transmittance, and absorption.

From Fig.~\ref{fig_conver}, it is clear that only a few terms in the sum of Eq.~\eqref{eq:Gb2D} are needed to achieve a good convergence. In this specific case, the contribution of higher orders is negligible beyond $|l|>2$. As a general recipe, it is necessary to sum all terms where $\mathrm{Re}\left[k_{||}^{(l)}\right] > 0$ plus two/three additional terms (this condition reduces to $\mathrm{Im}\left[k_{||}^{(l)}\right] = 0$ at real frequencies). In addition, we have estimated the value of $m^{(l)}$ as the number of terms needed in the sum to reduce the error in Eq.~\eqref{eq:Gb1D_l} below $10^{-3}$. This means that the correction of the $m^{(l)} + 1$ term is smaller than the millionth part of the total sum. Remarkable, only 9 terms are needed to converge to an accurate value to order $l = 0$; and  120 terms are enough for $|l|=1$. As $l$ is increased, $m^{(l)}$ becomes higher, but their contributions to the final result become negligible. Nonetheless, an increase of $m^{(l)}$ is expected with the decrease of the value of the sum, since more accuracy is necessary for the sum to converge. Alternatively, if $m^{(l)}$ were defined in relation to the total sum of all directive orders $l$ (with respect to the blue pillar in Fig.~\ref{fig_conver}), the value of $m^{(l)}$ would be smaller.

\subsection{2D arrays with arbitrary lattice symmetry}

For completeness, let us next consider a generic 2D array of particles defined by the lattice constants $a$ and $b$ with corresponding lattice vectors forming an angle $\theta$ between them. Then, if the primitive vector associated to the lattice constant $a$ is taken along the $x$ axis, the position of the particle labeled as $(n,m)$ becomes
\be
\r_{nm} = (na + m\cos\theta)\hat{x} + m\sin\theta\hat{y}.
\ee
For example, the rectangular lattice is recovered for $\theta = \pi/2$, while $\theta = \pi/3$ and $a=b$ yield a triangular lattice. Using this description the lattice depolarization dyadic, $\GGL_{b}$, follows the same expression shown in Eq.~\eqref{eq:Gb2D} but with some changes in the arguments of the functions:
\be
b \rightarrow b\sin\theta,
\ee
\be
k_y^{(0)} \rightarrow k_y^{(0)} + \dfrac{2\pi}{a}l\dfrac{\cos\theta}{\sin\theta}.
\ee 
Note that the expression that replaces $k_y^{(0)}$ also depends on the diffraction order index $l$. $\GGL_{b-Ch}$ does not change
\be
\GGL_{Ch}\left(a,k,k_x^{(0)}\right) \rightarrow \GGL_{Ch}\left(a,k,k_x^{(0)}\right),
\ee
while the arguments of $\GGL_{b-1D}$ are modified
\be
&& \GGL_{b-1D}^{(l)}\left(b,k,k_y^{(0)}, k_x^{(l)}\right) \rightarrow  \nonumber \\ &&
\hspace{1cm}\GGL_{b-1D}^{(l)}\left(b\sin\theta,k,k_y^{(0)} + \dfrac{2\pi}{a}l\dfrac{\cos\theta}{\sin\theta},k_x^{(l)}\right).
\ee
The same changes will affect the wavevector components of the diffracted orders. Specifically,
\be
k_y^{(p)} \rightarrow k_y^{(l,p)} =  k_y^{(0)} + \dfrac{2\pi}{a}l\dfrac{\cos\theta}{\sin\theta} - \dfrac{2\pi}{b\sin\theta}p,
\ee
while $k_x^{(l)}$ is unaffected. Thus, the tensor lattice sum components are modified as
\be
\GGL_{lp}^{\pm}e^{\pm\ii\phi_{lp}(\r,\theta)} =  \dfrac{\ii}{2ab\sin\theta k_{z}^{(lp)}} \mathbf{\mathcal{L}} e^{\pm\ii\phi_{lp}(\r,\theta)},
\ee
where we add the $\theta$ argument to the phase to account for the changes in the diffracted order wavevector.
%Alternatively, it is possible to define the chains of particles along the direction $\hat{u} = cos\theta \hat{x} + sin\theta \hat{y}$. 
%The expression for $\GGL_{b-1D}$ is shown in the Appendix??

\subsection{Complex unit cells}

In  general, the unit cell may in turn be composed of more than one particle. The self-interaction between the particles of the same kind is still described by $\GGL_b$, but it is necessary to also account for the interaction between the different particles within the unit cell. Interestingly, this interaction is easy to express in the reciprocal space through the tensor lattice sums by $\GGL_{ij} = \left[\GGL^{+}(\r_i - \r_j) + \GGL^{-}(\r_i - \r_j)\right]/2$, where $\r_{i}$ and $\r_{j}$ are the positions of particles $i$ and $j$ within the unit cell, respectively. Nonetheless, if two different particles are in the same $xy$ plane, the convergence is slow, but it can be improved using convergence techniques. 

\section{Resonances and bound states in the continuum in metasurfaces}
 
In order to find the resonant states of the metasurfaces we need to find solution to Eq.~\eqref{eq:cemd} in the absence of the external wave. At the condition $\Psi^{(0)} = 0$,  Eq.~\eqref{eq:cemd} becomes a homogeneous linear system of equations that supports solutions only when
\be
\left| \II - k^2\GGL_b\alphagg \right| = \left| \dfrac{1}{k^2\alphagg} - \GGL_b \right| = 0.
\label{eq:BS}
\ee
The complex frequencies at which Eq.~\eqref{eq:BS} is satisfied are the eigenfrequencies, denoted by $\nu$. In the latter eigenmode equation, it is more convenient 
%For the properties of $\GGL_{b}$ and $\alpha$ is better 
to use the second expression than the first one. Also, in this context, $k_x^{(0)}$ and $k_y^{(0)}$ are the in-plane wavevector components of the surface wave represented by the resonant mode, not related to any external wave. The eigenmode equations is written in the case of one particle per unit cell. For complex unit cells, $\alphagg$ is a matrix with the polarizability terms of all particles. In addition, $\GGL_b$ must be  replaced by a matrix that contains $\GGL_b$  and also  the interaction matrices $\GGL_{ij}$ (the specific form of these matrices to be determined by the basis chosen to describe the fields). Since $\GGL_{b}$ and $\GGL_{ij}$ can be expressed in the reciprocal space, Eq.~\eqref{eq:BS} can be thus employed to solve for the resonant modes of the metasurface. 

Below the light line (and below diffraction), the solutions of Eq.~\eqref{eq:BS} are real and determine the dispersion relation of guided modes that propagate along the metasurface. Opposite, above the light line the solutions are normally given by complex frequencies. The quality factor of the resulting mode is defined as the ratio between the real and the imaginary parts of the frequency, informing us of how fast the system leaks energy out to the continuum of radiation. However, it is also possible to find real solutions in this region, called bound states in the continuum (BICs)
~\cite{Abujetas2019c,Abujetas2019d}. Although they are embedded in the continuum of radiation, for symmetry reasons or interference effects these states remain localized within the metasurface without emitting energy to the far field \cite{Hsu2016a}.

\section{Practical case: square array of dielectric spheres}

Let us consider a specific case: a  square array of dielectric spheres of constant dielectric permittivity $\epsilon = 3.5$ (similar to that of Si in the visible and near-IR), with normalized lattice constant $a/R = b/R = 4$, where $R$ is the sphere radius. To describe the electromagnetic field, we choose the following basis:
\begin{equation}
 \bm{\Psi}(\r) = 
  \begin{bmatrix}
\E(\r) \\
Z\H(\r) 
\end{bmatrix},
\,
\E(\r) = 
\begin{bmatrix}
E_{x}(\r)\\
E_{y}(\r)\\
E_{z}(\r)
\end{bmatrix},
\,
\H(\r) = 
\begin{bmatrix}
H_{x}(\r)\\
H_{y}(\r)\\
H_{z}(\r)
\end{bmatrix},
\label{eq:EM_base}
\end{equation}
$\E(\r)$ and $\H(\r)$ being the electric and magnetic vector fields defined in Cartesian coordinates and  $Z=\left(\mu_0/\epsilon_0\right)^{1/2}$ the vacuum impedance. The linear differential operator $\mathbf{\mathcal{L}}$ takes the form
\be
\mathbf{\mathcal{L}} = 
\begin{pmatrix}
\left[ {\bf I} + \frac{\bm{\nabla} \bm{\nabla}}{k^2} \right] &  \ii \frac{\bm{\nabla}}{k} \times \ {\bf I} \\
- \ii \frac{\bm{\nabla}}{k} \times \ {\bf I} & \left[ {\bf I} + \frac{\bm{\nabla} \bm{\nabla}}{k^2} \right]
\end{pmatrix}.
\ee
%Appendix~\ref{} shows the tensor components of $\GGL_{b}$ in this electromagnetic base.

We assume that the sphere electrodynamic response can be fully described by its electric and magnetic dipolar contributions in terms of a polarizability tensor \cite{sersic2011magnetoelectric,albaladejo2010radiative}, $\alphagg$:
\begin{subequations}\begin{eqnarray}
&& \alphagg =
\begin{pmatrix}
\alphag^{(e)} & 0 \\ 
0 & \alphag^{(m)}
\end{pmatrix}, \\
&& k^2\alphag^{(e)} = \ii\dfrac{6\pi}{k} a_{1} \II, \quad k^2\alphag^{(m)} = \ii\dfrac{6\pi}{k}b_{1} \II,
\end{eqnarray}
\end{subequations}
where $\alphag^{(e)}$ and $\alphag^{(m)}$ are the electric and magnetic polarizabilities, and
 $a_{1}$ and $b_{1}$ are the (dimensionless) Mie coefficients~\cite{Bohren_Huffman}.
 
Within this description and for incidence along the $x$ axis, the specular reflectance, $R_{0}$, for both $p$- and $s$-polarized fields is:
%(se puede llevar a otro apendice)
\begin{subequations}
\begin{eqnarray}
&& R^{(p)}_0 = \left( \dfrac{k^{2}}{2kab\cos\theta_0} \right)^{2} \left| \gamma^{(p)}\Big(\widetilde{\alpha}_{y}^{(m)} +   \widetilde{\alpha}_{z}^{(e)}\sin^2\theta_0 \right.\nonumber\\&&
\left. +2k^{2} G_{byz}\widetilde{\alpha}_{y}^{(m)}\widetilde{\alpha}_{z}^{(e)}\sin\theta_0\Big) - \widetilde{\alpha}_{x}^{(e)}\cos^{2}\theta_0 \right|^{2},  \\ 
&& R^{(s)}_0 = \left( \dfrac{k^2}{2kab\cos\theta_0} \right)^{2} \left| \gamma^{(s)}\Big(\widetilde{\alpha}_{y}^{(e)} +   \widetilde{\alpha}_{z}^{(m)}\sin^2\theta_0 \right.\nonumber\\ &&\left. +2k^{2} G_{byz}\widetilde{\alpha}_{y}^{(e)}\widetilde{\alpha}_{z}^{(m)}\sin\theta_0\Big) -   \widetilde{\alpha}_{x}^{(m)}\cos^{2}\theta_0 \right|^{2},
\end{eqnarray}\label{eq:reflectance}
\end{subequations}
where $\theta_{0}$ is the angle of incidence ($k^{(0)}_z = k\cos\theta_0$, $k^{(0)}_x = k\sin\theta_0$), and the renormalized polarizability terms in  $ \widetilde{\alpha}$ are:
\begin{subequations}
\begin{eqnarray}
  k^{2}\widetilde{\alpha}_{i}^{(e)} = \left( \dfrac{1}{k^2\alpha^{(e)}} -G_{bii} \right)^{-1},  \\ 
  k^{2}\widetilde{\alpha}_{i}^{(m)} = \left( \dfrac{1}{k^2\alpha^{(m)}} -G_{bii} \right)^{-1},
\end{eqnarray}\label{eq:pol}
\end{subequations}
with 
\begin{subequations}
\begin{eqnarray}
  \gamma^{(p)} = \dfrac{1}{1 - k^{4} G_{bzx}^{2}\widetilde{\alpha}_{x}^{(m)}\widetilde{\alpha}_{z}^{(e)}},  \\ 
  \gamma^{(s)}  = \dfrac{1}{1 - k^{4}G_{bzx}^{2}\widetilde{\alpha}_{x}^{(e)}\widetilde{\alpha}_{z}^{(m)}},
\end{eqnarray} \label{eq:gamma}
\end{subequations}
$G_{bii}$ and $G_{bzx}$ being the matrix elements of $\GGL_{b}$. The polarization is defined in such away that $p$, (respectively, $s$) stands for TM (respectively, TE). It should be noted that this convention is the opposite of that in Ref.~\cite{Abujetas2018a}, where the "transverse" polarization was defined for the sake of convenience with respect to the cylinder axis, rather than with respect to the plane of incidence.
%(see Appendix~\ref{}).

On the other hand, Eq.~\eqref{eq:BS} can be used to find the resonant modes  supported by the metasurface. For those modes propagating along the $x$ axis ($k^{(0)}_x \not = 0$ and $k^{(0)}_y = 0$), although in this case the wavevector is not related to any incident external plane wave, Eq.~\eqref{eq:BS} reduces to
\be
\left|\eta^{(e)}_{x}\right| \times \left| \eta^{(m)}_{x}\right| \times \left|\eta^{(em)}_{yz}\right| \times  \left| \eta^{(me)}_{yz} \right| = 0,
\label{eq:eingen}
\ee
with 
\be
\eta^{(e)}_{x} = \dfrac{1}{k^2 \widetilde{\alpha}_{x}^{(e)} }, \quad  \eta^{(em)}_{yz} = \dfrac{1}{k^4 \widetilde{\alpha}_{y}^{(e)} \widetilde{\alpha}_{z}^{(m)}}  - G_{byz}^2, \nonumber \\
\eta^{(m)}_{x} = \dfrac{1}{k^2 \widetilde{\alpha}_{x}^{(m)} }, \quad  \eta^{(me)}_{yz} = \dfrac{1}{k^4 \widetilde{\alpha}_{y}^{(m)} \widetilde{\alpha}_{z}^{(e)} }  - G_{byz}^2.
\ee
Each term in Eq.~\eqref{eq:eingen} is associated to a different resonant surface mode. The solutions of $\eta^{(e)}_{x}$ and $\eta^{(m)}_{x}$ represent coherent oscillations of electric and magnetic dipoles along the $x$ axis, respectively. Since the imaginary part of $1/\widetilde{\alpha}^{(e,m)}_{x}$ %(and of $1/\widetilde{\alpha}^{(e,m)}_{y}$) 
(for lossless particles) vanishes only at $k^{(0)}_x \geq k$, they can never lead to a BIC; inside the continuum of radiation they represent broad resonant (leaky) modes that radiate energy to the far field. 
More interesting are the terms $\eta^{(em)}_{yz}$ and $\eta^{(me)}_{yz}$. They represent hybrid modes where electric (magnetic) dipoles along the $y$ axis are coupled with magnetic (electric) dipoles along the $z$ axis. These terms can yield BICs due to the mutual interference between the different dipolar modes, although in general they correspond to broad resonant modes. Nonetheless, at $k^{(0)}_x = 0$, both $G_{byz} = 0$ and the imaginary part of $1/\widetilde{\alpha}_{z}$ are zero. Then, the dipolar modes are decoupled and the metasurface can support a BIC given by the in-phase oscillation of dipoles along the $z$ axis. This BIC is the typical one used in several applications \cite{Kodigala2017,Ha2018}. In addition, all terms can support guided modes in the region defined by $k^{(0)}_x \geq k$ and $k^{(0)}_x \leq 2\pi/a - k$ (with $k \geq 0$, and $k^{(0)}_x \geq 0$), where the imaginary part (for lossless particles) of Eq.~\eqref{eq:eingen} is identically zero.

\begin{figure}
\includegraphics[width=1\columnwidth]{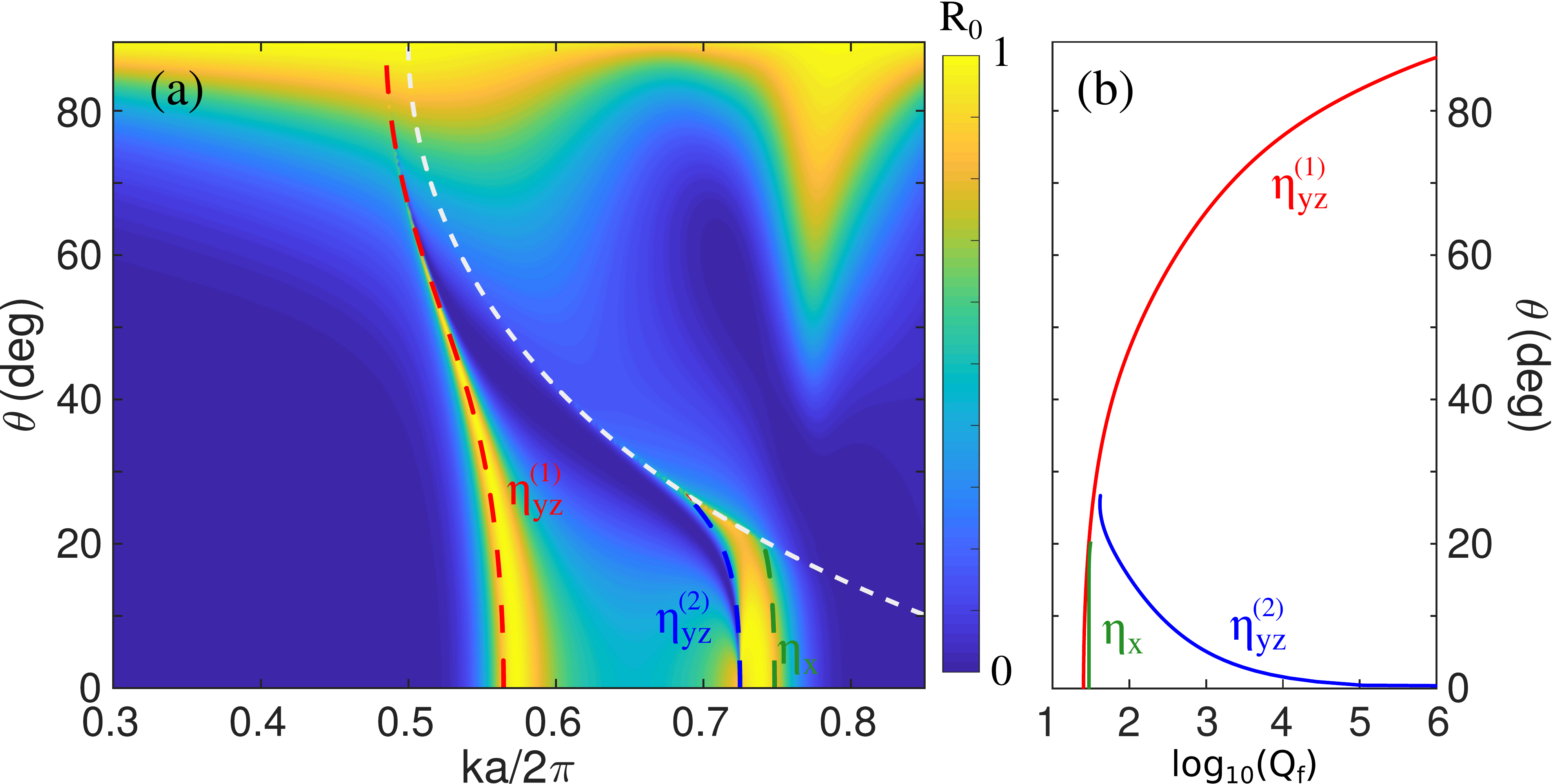}
\caption{(a) Color map of the reflectance for TM polarization for a square array of dielectric spheres of constant dielectric permittivity $\epsilon = 3.5$ (similar to that of Si in the visible and near-IR), with normalized lattice constant $a/R = b/R = 4$, where $R$ is the sphere radius. Dispersion relations of the resonant surface modes given by the zeros of $\eta^{(e)}_{x}$ and $\eta^{(me)}_{yz}$ are superimposed as dashed color lines, while the white dashed lines delimit the diffractive region. (b) Q-factor of the surface modes.}
\label{fig_RyBandas_TM}
\end{figure}

To show the interplay of the resonant mode on the properties of the metasurface we study the specular reflectance, $R_0$, for both polarizations. First, in Fig.~\ref{fig_RyBandas_TM}a) the specular reflectance for TM polarized light and the dispersion relation of the resonant modes are shown as a function of the normalized frequency and angle of incidence. Due to the symmetry of the resonant modes not all of them can be exited at a given polarization (keep in mind that $k^{(0)}_y = 0$), so we only show the modes related to the zeros of $\eta^{(e)}_{x}$ and $\eta^{(me)}_{yz}$. Since the incident polarization determines the nature of the mode (electric or magnetic), for the sake of clarity the superscript is replaced by a number that label the mode solution. As expected, the resonances of the reflectance coincide with dispersion relation of the surface modes. In addition, their widths agree with the Q-factor of the modes, displaying in Fig.~\ref{fig_RyBandas_TM}b). The Q-factor is calculated as the ratio between the real and imaginary part of the eigenfrequency, $\nu = \nu' + \ii \nu''$. For simplicity we only represent the modes in the non-diffracting region (delimited by the white dashed line), but Eq.~\eqref{eq:eingen} still have solutions in this region.   

Around $ka/(2\pi) = 0.56$ there is the $\eta^{(1)}_{yz}$ resonant (leaky) surface mode. Its associate resonance in the reflectance spectra is broad at normal incidence, and it becomes narrower as long as the angle of incidence increases. The Q-factor of the leaky surface mode diverges at $k^{(0)}_x = k$, point where the leaky mode becomes a guided mode. Later, as the frequency increases in the reflectance we find a narrow resonance around $ka/(2\pi) = 0.72$, related to the $\eta^{(2)}_{yz}$ surface mode. At normal incidence this mode represents a symmetry protected BIC (in-phase oscillation of electric dipoles along the $z$ axis) that becomes broader as the angle of incidence increases. However, due to diffraction, the dispersion relation of the modes stops at the diffraction line. Finally, the last surface mode encloses in the studied frequency windows is $\eta^{(1)}_{x}$. The width of the surface resonance in almost constant as a function of the angle of incidence and as before its dispersion relation stops at diffraction.

\begin{figure}
\includegraphics[width=1\columnwidth]{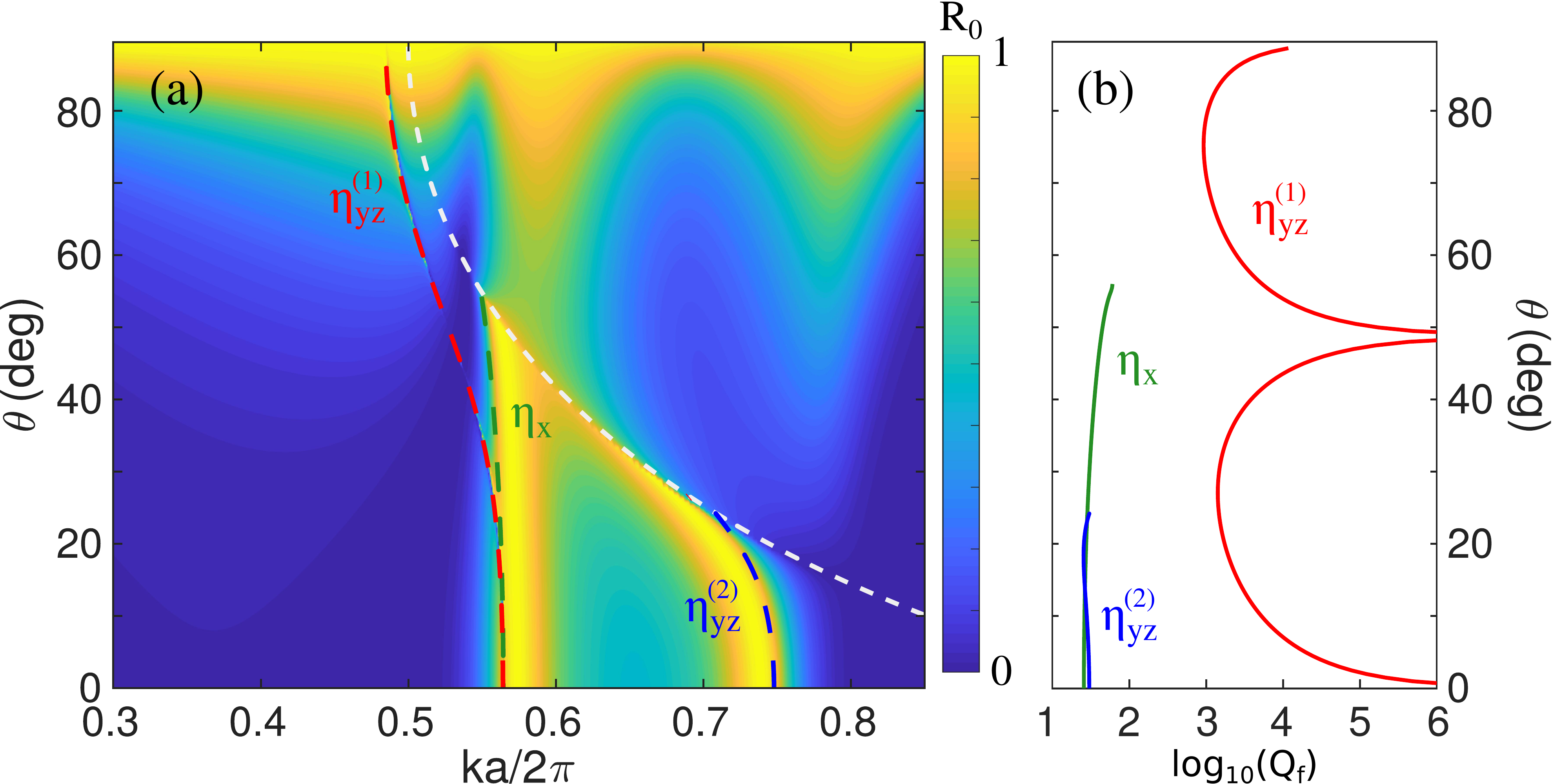}
\caption{(a) Color map of the reflectance for TE polarization for a square array of dielectric spheres of constant dielectric permittivity $\epsilon = 3.5$ (similar to that of Si in the visible and near-IR), with normalized lattice constant $a/R = b/R = 4$, where $R$ is the sphere radius. Dispersion relations of the resonant surface modes given by the zeros of $\eta^{(m)}_{x}$ and $\eta^{(em)}_{yz}$ are superimposed as dashed color lines, while the white dashed lines delimit the diffractive region. (b) Q-factor of the surface modes.}
\label{fig_RyBandas_TE}
\end{figure}

The agreement between the dispersion relation of the surface resonant modes and the resonances in the reflectance spectra is further confirmed for TE incident waves, as it can be seen in Fig.~\ref{fig_RyBandas_TE}. Now, for this polarization $\eta^{(m)}_{x}$ and $\eta^{(em)}_{yz}$ are the relevant modes that can be excited by the incident external wave. Around $ka/(2\pi) = 0.57$, at normal incidence there are two modes, $\eta^{(1)}_{x}$ and $\eta^{(1)}_{yz}$. The characteristics of $\eta^{(1)}_{x}$ are similar that the ones observed in TM polarization for its electric counterpart. The mode is relatively broad and presents low dispersion in frequency that ends up at the diffraction line. Opposite, $\eta^{(1)}_{yz}$ posses more interesting features. First, at normal incidence it represents a symmetry protected BIC given by in-phase oscillation of magnetic dipoles along the $z$ axis. As $k_x^{(0)}$ increases the surface mode becomes broader and evolves until a minimum in the Q-factor. From this point, the mode becomes narrower again and the Q-factor finally diverges around $\theta = 48^{\circ}$; the surface modes turns into an accidental BIC given by the destructive interference at the far field between the emission of the electric dipoles along the $y$ axis and the magnetic dipoles along the $z$ axis. Lastly, the Q-factor decreases to another minimum and diverges once again at $\theta = 90^{\circ}$, where the surface mode becomes a guided mode. For the sake of completeness, the last surface mode shown in Fig.~\ref{fig_RyBandas_TE} is $\eta^{(2)}_{yz}$, placed around $ka/(2\pi) = 0.75$. This mode is much broader than $\eta^{(1)}_{yz}$, although both modes come from to the same term ($\eta^{(em)}_{yz}$). Also, its dispersion relation stops at the diffraction line. 

\begin{figure}
\includegraphics[width=1\columnwidth]{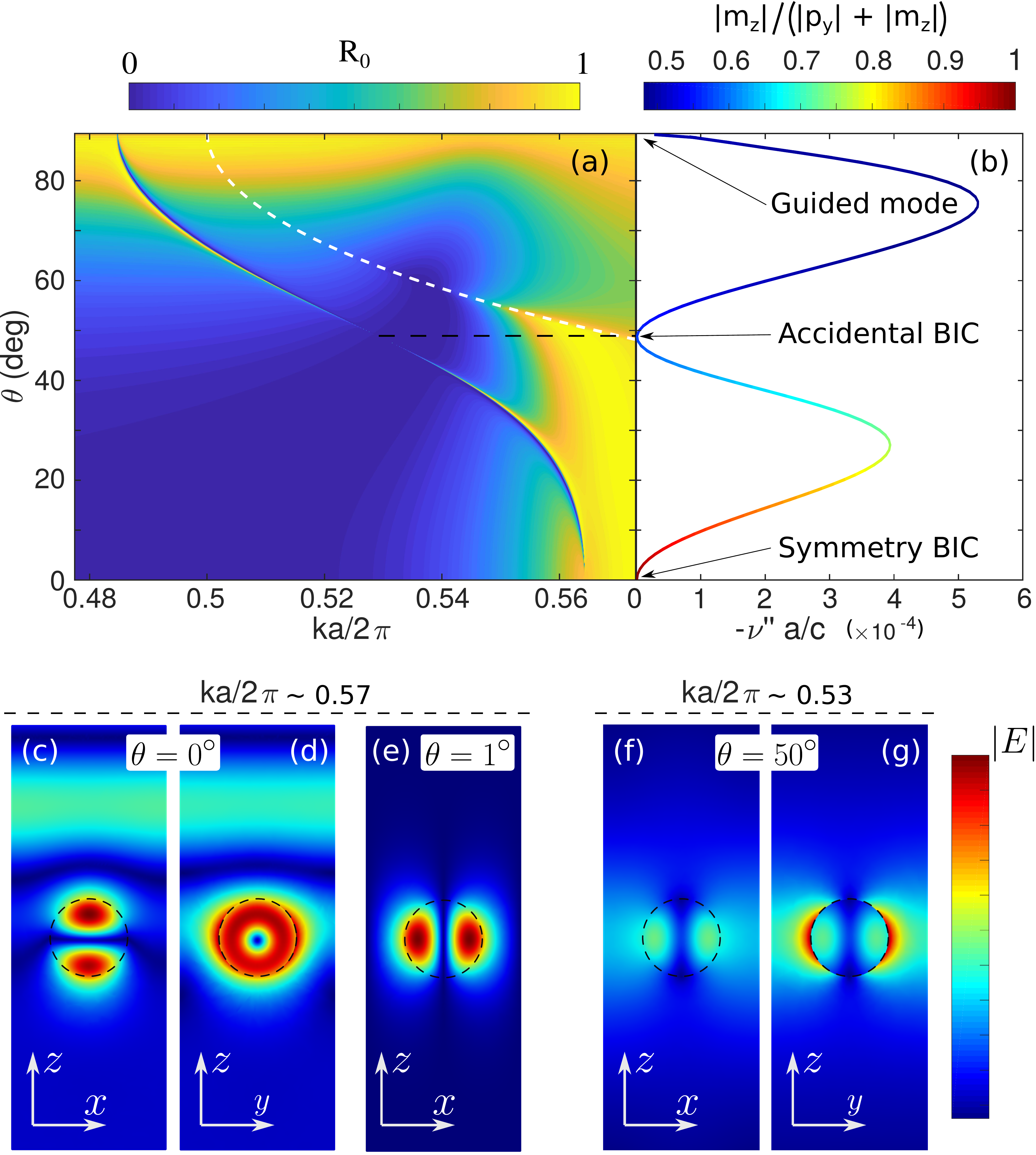}
\caption{(a) Zoom to the reflectance for TE polarization from Fig.~\ref{fig_RyBandas_TE} in the region of the accidental BIC. The white dashed line delimits the diffractive region. (b) Imaginary part of the eigenvalue of the narrow $\eta^{(em)}_{yz}$ mode.
%and chracterization of the eigenvector. 
(c-g) Near field distributions at (c-e) $ka/(2\pi) \sim 0.57$ (symmetry-protected BIC) and (f-g) $ka/(2\pi) \sim 0.53$ (accidental BIC) for different angles of incidence. The incident wave goes from top to bottom and the plane of incidence is the $xz$ plane.}
\label{fig_ZoomBic}
\end{figure}

The mode $\eta^{(1)}_{yz}$ that comes from $\eta^{(em)}_{yz}$ is very narrow in all regions, so its features on the reflectance spectra Fig.~\ref{fig_RyBandas_TE}a) are hidden by the dashed line that marks the dispersion relation. Thus, let us take a closer view of this region in Fig.~\ref{fig_ZoomBic}a), where a zoom on the reflectance for TE polarization is done. Strong asymmetry Fano resonances in the reflectance can be appreciated throughout all angles of incidence. The asymmetry of the resonance will depend on the interference given by the rest of surface modes present in the metasurface. Also, the resonance shows vanishing features distinctive of BIC at normal incidence, around $\theta = 48^{\circ}$ and at grazing incidence, that corresponds to the symmetry protected BIC, the accidental BIC and the connection to the guided mode, respectively.

In order to characterize the surface hybrid mode, Fig.~\ref{fig_ZoomBic}b) shows the imaginary part of the eigenvalue (inversely proportional to the Q-factor), whose zeros coincide with the BICs (and with the guided mode). In addition, the color of the line represents the eigenvector of the mode. The red color means that the mode is made purely of magnetic dipoles along the $z$ axis, while the line turns into blue as long as the nature of the mode becomes more hybrid. As predicted, the eigenvalue of the symmetry protected BIC has no $p_y$ component, it is formed only by the oscillation of magnetic $m_z$ dipoles. Contrary, the accidental BIC is a hybrid mode, where the mutual interference between the emission of $p_y$ and $m_z$ cancels out at that specific configuration.

Narrow resonances in the reflectance spectra are normally associated with large enhancements in the near field distribution. As a representative case, we show in Fig.~\ref{fig_ZoomBic}c-e  the norm of the electric field in one unit cell at the frequency of the symmetry-protected BIC, calculated through numerical methods (COMSOL). At normal incidence, the coupling to the BIC is forbidden and the near field is dominated by the resonance of the magnetic dipoles along the $x$ axis ($\eta^{(m)}_{x}$), as can be seen in the different plane cuts, where the fields present a minimum along the $x$ axis. The maximum of the near electromagnetic field is about three times higher than the incident external field, in concordance with the relative small quality factor of the corresponding resonance, $Q_f > 10^2$. Notably, when the angle of incidence is slightly modified, the field distribution entirely changes. Now, the small field component of the incident wave perpendicular to the metasurface couples efficiently with the narrow resonance related to the quasi-BIC. The field distribution is mainly given by the oscillation of magnetic dipoles along the $z$ axis, and the near field is enhanced by several orders of magnitude (above 10$^3$). In fact, due to the huge enhancement at the quasi-BIC, the wave scattered to the far field cannot be appreciated in Fig.~\ref{fig_ZoomBic}e, unlike that in Fig.~\ref{fig_ZoomBic}c-d. As a final remark, recall that the field enhancement diverges together with the Q-factor as the angle of incidence goes to zero, although exactly at normal incidence there is no plane wave coupling to the BIC as mentioned above; in practice, inherent losses of real materials will prevent this divergence. Note that due to the high symmetry along the $z$ axis, the fields at the $yz$ plane (not shown here) are very similar to those at the $xz$ plane.

When the angle of incidence increases, the nature of the resonant $\eta^{(em)}_{yz}$ mode becomes hybrid, and this can be also observed in the near field. In Fig.~\ref{fig_ZoomBic}f-g, the near field at the resonant condition is shown for $\theta = 50^{\circ}$, slightly above (in angles) the accidental BIC condition ($\theta_{a-BIC} \sim 48^{\circ}$). The electromagnetic field presents features of both contributions: namely, magnetic dipoles along the $z$ axis and electric dipoles along the $y$ axis. Also, since the field is calculated at the resonance of the reflectance, the field enhancement is huge, close to 10$^3$. 

\section{Conclusions}
We have developed a coupled electric and magnetic dipole (CEMD) analytical formulation to describe the reflection and transmission of planar arrays of electric and/or magnetic particles, including specular and diffractive orders, and modes emerging as poles of such lattice Green function can be extracted. The formulation is largely simplified by rewriting the 2D lattice Green function in terms of a 1D (chain) version that fully converges.  Electric and/or magnetic dipoles with all three orientations arising in turn from a single or various meta-atoms per unit cell are considered. Analytical expressions for the emergence of either guided or resonant modes are given, along with corresponding Q-factors, which allow us to identify bound states in the continuum. By way of example, both symmetry-protected and accidental BICs are identified for a square array of all-dielectric spheres with constant dielectric permittivity close to that of Si in the visible and near-IR.
Despite limited to dipolar meta-atoms, this formulation can be exploited to deal with metasurfaces/metagratings of interest throughout the electromagnetic spectrum, bearing in mind that  most plasmonic, all-dielectric, and/or hybrid meta-atoms, either at or out of resonance, behave in many spectral regimes as a combination of electric and magnetic dipoles.  In this regard, future work incorporating higher complexity in the optical response of the meta-atoms such as core-shell particles and/or higher-order multipoles \cite{Paniagua-Dominguez2011,RaDi2013,Sheverdin2019} will no doubt enrich the phenomenology.

\begin{acknowledgments}
The authors dedicate this work to the memory of their beloved colleague and friend, Prof. Juan José Sáenz, who passed away on March 22, 2020.
This work has been supported by the Spanish Ministerio de Ciencia e Innovación  (MICIU/AEI/FEDER, UE) through the grants MELODIA (PGC2018-095777-B-C21) and NANOTOPO (FIS2017-91413-EXP), and from the Ministerio de Educación, Cultura y Deporte through a PhD Fellowship (FPU15/03566).  D.R.A. and J.A.S.G. acknowledge fruitful discussions with R. Paniagua-Domínguez.
\end{acknowledgments}

\appendix

\section{Lattice depolarization dyadic for an array of cylinders}
When the electromagnetic field is described as shown in Eq.~\eqref{eq:EM_base},
the lattice depolarization dyadic for an array of cylinders (with their axes along the $y$ axis), $\GGL_{b-1D}$, can be written as
\be
\GGL_{b-1D}^{(l)}(b,k,k_y,k_x) = \begin{pmatrix} \GL_{b-1D} & \GL^{(EM)}_{b-1D} \\ -  \GL^{(EM)}_{b-1D} & \GL_{b-1D}
\end{pmatrix},
\ee 
where the matrix terms of $\GL_{b-1D}$ and $\GL^{(EM)}_{b-1D}$ are:
\begin{widetext}
\begin{align}
(\GL_{b-1D})_{xx} &= \dfrac{k_{||}^2}{k^2}\left( \ii\left[\dfrac{1}{2q_{0}b} -\dfrac{1}{4} \right] + \dfrac{1}{2b}\sum\limits_{m=1}^{\infty} \left( \dfrac{\ii}{q_{m}} + \dfrac{\ii}{q_{-m}} - \dfrac{2}{k_{m}}\right) + \dfrac{1}{2\pi}\left[\ln\dfrac{k_{||}b}{4\pi} + \gamma_{e} \right] \right), \\
 (\GL_{b-1D})_{yy} &= \ii\left[ \dfrac{q_0^2 + k_x^2}{2k^2q_0 b} - \dfrac{1}{8}\dfrac{k^2 + k_x^2}{k^2} \right] + \dfrac{1}{4\pi}\left[\ln\dfrac{k_{||}b}{4\pi} + \gamma_{e} \right]\dfrac{k^2 + k_x^2}{k^2} + \dfrac{1}{8}\dfrac{K_0^2 - q_0^2}{k^2} + \dfrac{\pi}{6k^2b^2}  \nonumber \\
 & + \dfrac{1}{2k^{2}b}\sum\limits_{m=1}^{\infty} \left( \ii \dfrac{k^2 - K_{m}^2}{q_{m}} + \ii \dfrac{k^2 - K_{-m}^2}{q_{-m}} - \dfrac{k^2 + k_x^2 - 2k_m^2}{k_{m}}\right), \\
 (\GL_{b-1D})_{zz} &= \ii\left[ \dfrac{K_0^2 + k_x^2}{2k^2q_0 b} - \dfrac{1}{8}\dfrac{k^2 + k_x^2}{k^2} \right] + \dfrac{1}{4\pi}\left[\ln\dfrac{k_{||}b}{4\pi} + \gamma_{e} \right]\dfrac{k^2 + k_x^2}{k^2} - \dfrac{1}{8}\dfrac{K_0^2 - q_0^2}{k^2} - \dfrac{\pi}{6k^2b^2}  \nonumber \\
 & + \dfrac{1}{2k^{2}b}\sum\limits_{m=1}^{\infty} \left( \ii \dfrac{k^2 - q_{m}^2}{q_{m}} + \ii \dfrac{k^2 - q_{-m}^2}{q_{-m}} - \dfrac{k^2 + k_x^2 + 2k_m^2}{k_{m}}\right), \\
\left(\GL_{b-1D}\right)_{xy} & = - \dfrac{k_x}{k}\left(\ii \frac{K_0}{ 2 k q_0 b} +  \frac{\ii}{2 k b} \sum_{m=1}^{\infty} \left( \dfrac{K_m}{q_{m}} +  \dfrac{K_{-m}}{q_{-m}} \right) - \dfrac{1}{2\pi}\dfrac{K_{0}}{k} \right), \\
\left(\GL^{(EM)}_{b-1D}\right)_{yz} & = - \left(\GL^{(EM)}_{b-1D}\right)_{zy}   \nonumber \\ 
&=
  \dfrac{k_x}{k}\left( \ii\left[\dfrac{1}{2q_{0}b} -\dfrac{1}{4} \right] \right. + \left. \dfrac{1}{2b}\sum\limits_{m=1}^{\infty} \left( \dfrac{\ii}{q_{m}} + \dfrac{\ii}{q_{-m}} - \dfrac{2}{k_{m}}\right) + \dfrac{1}{2\pi}\left[\ln\dfrac{k_{||}b}{4\pi} + \gamma_{e} \right] \right), \\
%  \label{eq:gbEMyz}
\left(\GL^{(EM)}_{b-1D}\right)_{zx} &= - \left(\GL^{(EM)}_{b-1D}\right)_{xz} =
  \ii \frac{K_0}{ 2 k q_0 b} +  \frac{\ii}{2 k b} \sum_{m=1}^{\infty} \left( \dfrac{K_m}{q_{m}} +  \dfrac{K_{-m}}{q_{-m}} \right) - \dfrac{1}{2\pi}\dfrac{K_{0}}{k}.
  \label{eq:gbEM1D}
 \end{align} 
\end{widetext}
$b$ is the distance between cylinders and
\be
k_m = \frac{2 m \pi}{b}, \qquad K_m \equiv k_y - k_m, \nonumber \\ q_m = \sqrt{ k_{||}^2 - K_m^2}, \qquad k_{||}^2 = k^2 - k_x^2. 
%\label{eq:ks}.
\ee
The rest of the matrix elements are equal to zero.
For $k_x = 0$ (incidence perpendicular to the cylinder) we recover the expressions given in Ref~\cite{Abujetas2018a}.

\section{Lattice depolarization dyadic for a chain of particles}

Using the same description for the electromagnetic field as in Appendix A, the lattice depolarization function of a chain of particles (placed along the $x$ axis), $\GGL_{b-Ch}$, is written as
\be
\GGL_{b-Ch}(a,k,k_{x}^{(0)}) = \begin{pmatrix} \GL_{b-Ch} & \GL^{(EM)}_{b-Ch} \\ -  \GL^{(EM)}_{b-Ch} & \GL_{b-Ch}
\end{pmatrix},
\ee 
and the matrix elements of $\GL_{b-Ch}$ and $\GL^{(EM)}_{b-Ch}$ are: 
\begin{widetext}
\begin{align}
(\GL_{b-Ch})_{xx} & = -\dfrac{\ii}{ 2a^3k^2\pi}\left\lbrace \right. ak\left[ Li_{2}\left(e^{\ii(k-k_x)a}\right) + Li_{2}\left(e^{\ii(k+k_x)a}\right) \right]  \nonumber \\
& \qquad \qquad \qquad \left. +  \ii \left[ Li_{3}\left(e^{\ii(k-k_x)a}\right) + Li_{3}\left(e^{\ii(k+k_x)a}\right) \right]\right\rbrace, \\
(\GL_{b-Ch})_{yy} & = -\dfrac{\ii}{4a^3k^2\pi}\left\lbrace  \right.  -\ii a^2k^2\left[ Li_{1}\left(e^{\ii(k-k_x)a}\right) + Li_{1}\left(e^{\ii(k+k_x)a}\right) \right]  \nonumber \\
& \qquad \qquad \qquad + ak\left[ Li_{2}\left(e^{\ii(k-k_x)a}\right) + Li_{2}\left(e^{\ii(k+k_x)a}\right) \right] \nonumber \\
& \qquad \qquad \qquad \left. +  \ii \left[ Li_{3}\left(e^{\ii(k-k_x)a}\right) + Li_{3}\left(e^{\ii(k+k_x)a}\right) \right]\right\rbrace, \\
(\GL_{b-Ch})_{zz} & = (\GL_{b-Ch})_{yy}, \\
(\GL_{b-Ch}^{(EM)})_{yz} & = - (\GL_{b-Ch}^{(EM)})_{zy} = \dfrac{\ii}{4a^3k^2\pi}\left\lbrace  \right.  -\ii a^2k^2\left[ Li_{1}\left(e^{\ii(k-k_x)a}\right) - Li_{1}\left(e^{\ii(k+k_x)a}\right) \right] \nonumber \\
& \qquad \qquad \qquad \qquad \qquad \qquad \left. + ak\left[ Li_{2}\left(e^{\ii(k-k_x)a}\right) - Li_{2}\left(e^{\ii(k+k_x)a}\right) \right] \right\rbrace .
\end{align}
\end{widetext}
$a$ is the distance between particles and $Li_{s}(z)$ is the polylogarithm function, also known as Jonqui\'ere's function. For the special case of $s=1$, the polylogarithm function is $Li_{1}(z) = -ln(1-z)$. The rest of the matrix elements are equal to zero.

\bibliography{library}% Produces the bibliography via BibTeX.

\end{document}